\documentclass[12pt,preprint]{aastex}
\usepackage{natbib}
\bibpunct{(}{)}{;}{a}{}{,}
\shorttitle{FUV diffuse emission from the SMC}
\shortauthors{Pradhan, Pathak \& Murthy}

\begin{document}

\title{Observations of Far-Ultraviolet Diffuse Emission from the Small Magellanic Cloud}
\author{Ananta C. Pradhan\thanks{Email: acp@iiap.res.in}, Jayant Murthy$^{1}$ and Amit Pathak$^{2}$\\
$^{1}$ Indian Institute of Astrophysics, Bangalore - 560 034, India\\
$^{2}$ Department of Physics, Tezpur University, Tezpur - 784 028, India}

\begin{abstract}

We report the first observations of far-ultraviolet (FUV: 1000 -- 1150 \AA) diffuse radiation from the Small Magellanic Cloud (SMC) using observations from the {\em Far Ultraviolet Spectroscopic Explorer (FUSE)}. The strength of FUV diffuse surface brightness in the SMC ranges from the detection limit of 2000 photons cm$^{-2}$ s$^{-1}$ sr$^{-1}$ \AA$^{-1}$ to a maximum of $3 \times 10^{5}$ photons cm$^{-2}$ s$^{-1}$ sr$^{-1}$ \AA$^{-1}$ at 1004 \AA. The contribution of diffuse emission to the total radiation field was found to be 34\% at 1004 \AA\ to 44\% at 1117 \AA\ with a maximum observed uncertainty of 30\%. There is a striking difference between the FUV diffuse fraction from the SMC and the Large Magellanic Cloud (LMC) with the SMC fraction being higher probable because the higher dust albedo. The FUV diffuse emission correlates with H$\alpha$ emission in the H {\small II} regions of the SMC.

\end{abstract} 

\section{Introduction}

The Small Magellanic Cloud (SMC) is a nearby dwarf galaxy \citep[$\approx$ 60 Kpc;][]{Hilditch05} which provides an ideal environment to study the interstellar medium (ISM) in a region of low metallicity \citep[Z $\approx$ 0.005;][]{Dufour84, Asplund04}. The foreground Galactic extinction is low \citep[E(B-V) = 0.02 mag;][]{Hutchings82} and its face-on view orientation  allows the observer to investigate the small scale structures. The SMC, itself, contains significant amount of dust and gas but with a low dust-to-gas ratio \citep[8 times smaller than the Milky Way;][]{Bouchet85}, and a strong interstellar ultraviolet (UV) radiation field \citep[4 -- 10 times higher than that in the solar neighborhood;][]{Vangioni-Flam80}. The ISM of the SMC is similar to that of high redshift galaxies because of its low metallicity and therefore may be a stepping stone to our understanding of the ISM in them \citep{Witt00}. Dust in the SMC is quite different from that in either the Milky Way or the LMC as shown, for instance, by the absence of 2175 \AA\ bump \citep{Gordon03}. Models of the dust in the SMC typically assume that it is dominated by silicates with the absence of the 2175 \AA\ bump attributed to a lack of carbonaceous dust \citep{Weingartner01}.

The surface brightness and integrated magnitudes of the bright regions of the SMC has been mapped in the near-ultraviolet (NUV) by a number of rocket and satellite observations \citep{Nandy78, Vangioni-Flam80, Maucherat-Joubert80, Cornett94}. Here, we present the first observations of diffuse far-UV (FUV: 1000 -- 1150 \AA) emission from the SMC. These were serendipitous observations made with the {\em Far Ultraviolet Spectroscopic Explorer (FUSE)} and include different environments in the SMC: from those near hot stars to those further out in the edges of the galaxy. The diffuse emission tracks the interaction of the radiation field with the dust and is an important input into models of distant galaxies \citep{da Cunha08}. The SMC offers an opportunity to test these models at high spatial resolution and to distinguish the different components of the galaxy. 

\section{Observations and Data Analysis}

We have used observations made by the {\em FUSE} spacecraft to measure the diffuse emission from the SMC in the FUV. The {\em FUSE} instrument and its mission have been discussed by \citet{Moos00} and \citet{Sahnow00}. It consisted of four optical channels with each channel comprising a mirror, focal plane assembly (FPA) and diffraction grating. Two of the channels included optics coated with LiF and aluminum and the other two with SiC and each channel was imaged onto a delay-line detector at the focal plane. Observations were made through three different apertures: the high-resolution aperture (HIRS: \(1''.25\) $\times$ \(20''\)); the medium-resolution aperture (MDRS: \(4''\) $\times$ \(20''\)); and the low-resolution aperture (LWRS: \(30''\) $\times$ \(30''\)), with all three obtaining data simultaneously. Thus even though a source may have been observed in the MDRS or HIRS aperture, the diffuse background could still be measured through the LWRS aperture as it is separated from the former apertures by \(100''\) and \(200''\) respectively. Only the very brightest backgrounds could be observed with the smaller MDRS aperture or with the less sensitive SiC channels. \citet{Murthy04} have shown that the practical limit for {\em FUSE} diffuse observation is about 2000 photons cm$^{-2}$ s$^{-1}$ sr$^{-1}$ \AA$^{-1}$.

There were a total of 220 {\em FUSE} observations within $5^{\circ}$ of the SMC but 190 were of stars through the LWRS aperture leaving 30 pointings from which we could extract the diffuse background. These observations were from two classes of targets: stars observed through either of the MDRS or HIRS apertures; or empty areas of the sky where the spectrographs were allowed to thermalize before an instrumental realignment. The observational details of these targets are given in Table \ref{tbl-1}. Most of the regions observed are either active areas of star formation or H {\small II} regions, such as NGC 346 and NGC 330.

The data selection and analysis procedure have been explained in detail elsewhere \citep{Murthy04}. We began with the raw photon list and processed the data through the latest version of CalFUSE \citep[v3.2;][]{Dixon07} except that we estimated the instrumental background from the counts in the detector just off the spectrum. The background was subtracted from the data which was then collapsed into two wavelength bands per segment, avoiding airglow lines. This resulted in a total of six bands from three segments. We found that the data were of much higher quality from segment 1 leaving us with four bands at effective wavelengths of 1004 \AA\ (1A1), 1058 \AA\ (1A2), 1117 \AA\ (1B1), 1157 \AA\ (1B2).

The surface brightness measured in the {\em FUSE} bands show a strong correlation between each other with correlation coefficients of better than 0.9. Our observed surface brightnesses (Table \ref{tbl-1}) range from near the {\em FUSE} detection limit to as high as $3 \times 10^{5}$ photons cm$^{-2}$ s$^{-1}$ sr$^{-1}$ \AA$^{-1}$ in NGC 346, the youngest and largest H {\small II} region in the SMC. We have estimated the level of Galactic background at these wavelengths from the {\em Voyager} maps of \citet{Murthy99} to be about 1000 photons cm$^{-2}$ s$^{-1}$ sr$^{-1}$ \AA$^{-1}$, much less than the observed SMC fluxes.

\section {Results and Discussion} 

\begin{figure}
\plotone{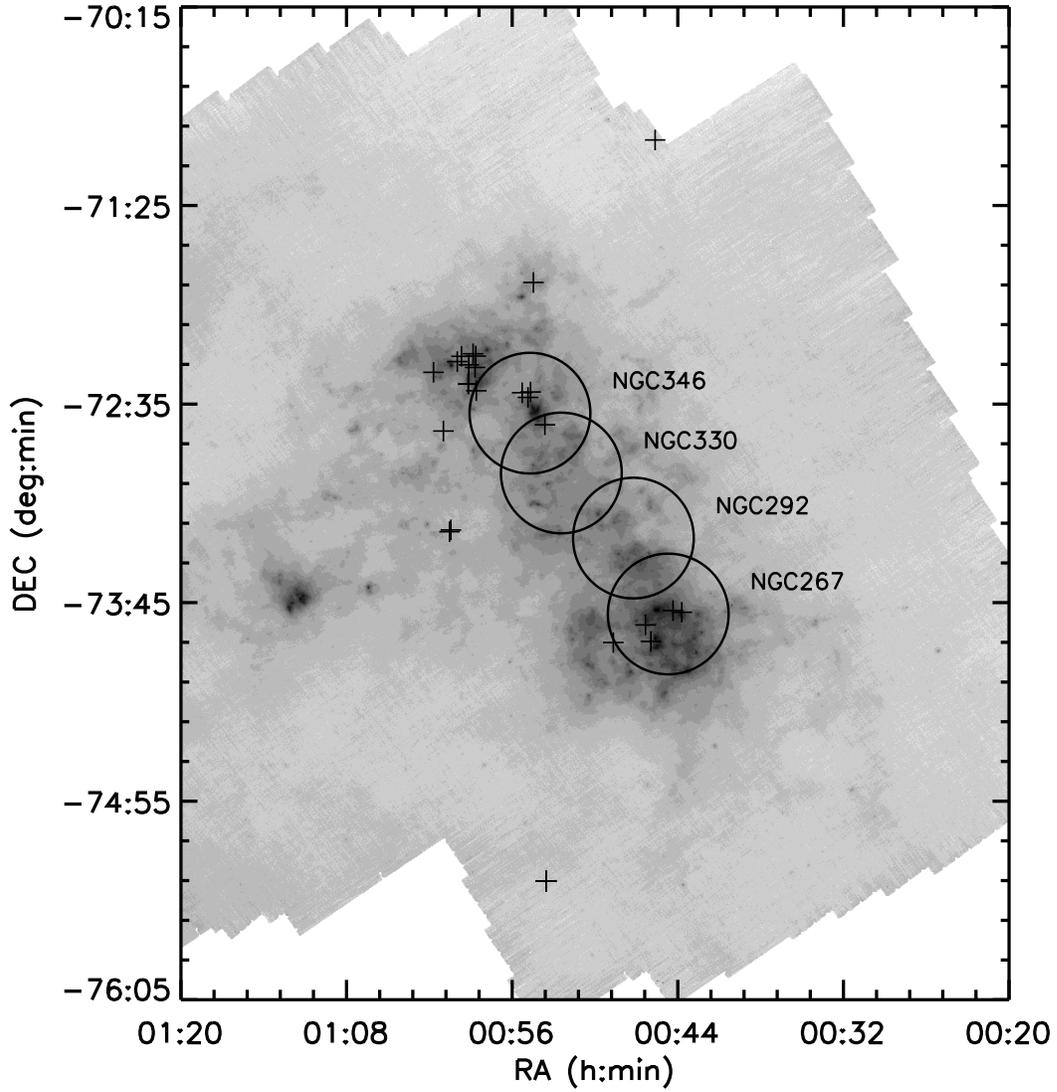}
\caption{IR 160 $\mu$m image of the SMC from \citet{Gordon09} showing the position of the diffuse {\em FUSE} targets marked by `$+$' symbols. \(40'\) diameter {\em UIT} observations have been shown by circles.
\label{Fig1}}
\end{figure}

\begin{figure}
\plotone{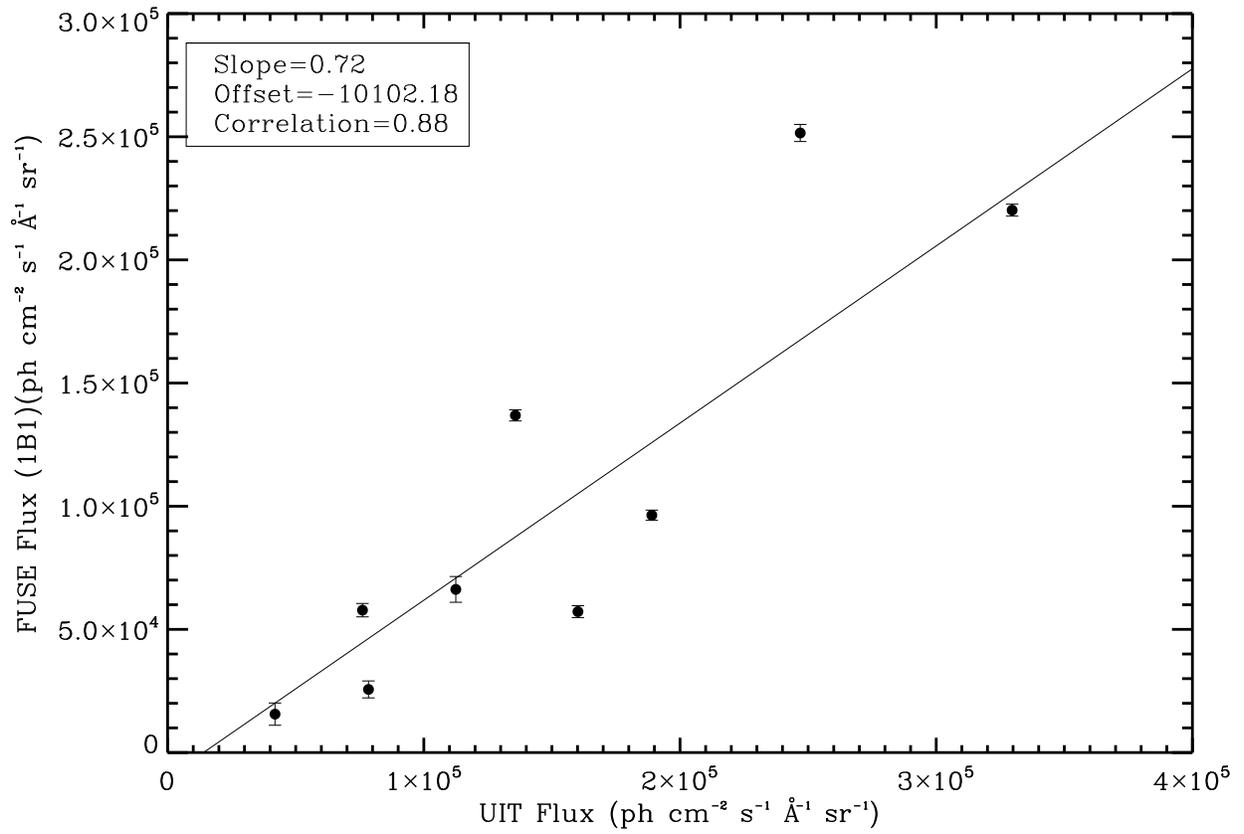}
\caption{Correlation between the {\em FUSE} (1B1) and the {\em UIT} surface brightness is shown. The correlation coefficient is 0.88. The best fit line is with slope 0.72 and an offset of -10102.18 photons cm$^{-2}$ s$^{-1}$ sr$^{-1}$ \AA$^{-1}$. Errors in the {\em FUSE} observations are shown by vertical bars and the error in the {\em UIT} observations is 15\%.
\label{Fig2}}
\end{figure}

We have plotted the location of our targets (plus symbols) on a 160 $\mu$m image of the SMC \citep{Gordon09} in Figure \ref{Fig1}. Also shown are the \(40'\) fields (circles) observed by {\em Ultraviolet Imaging Telescope (UIT)} at 1615 \AA, covering most of the SMC Bar \citep{Cornett97}. We calculated the diffuse NUV flux for the 9 {\em FUSE} locations that are within the {\em UIT} field of observations by integrating the \(1''.13\) {\em UIT} pixels over the \(30''\) $\times$ \(30''\) {\em FUSE} LWRS aperture. These fluxes are listed in Table \ref{tbl-1}. The {\em UIT} fluxes are highly correlated with the surface brightness of the {\em FUSE} bands with a correlation coefficient of better than 0.88 (Figure \ref{Fig2}).

\begin{figure}
\plotone{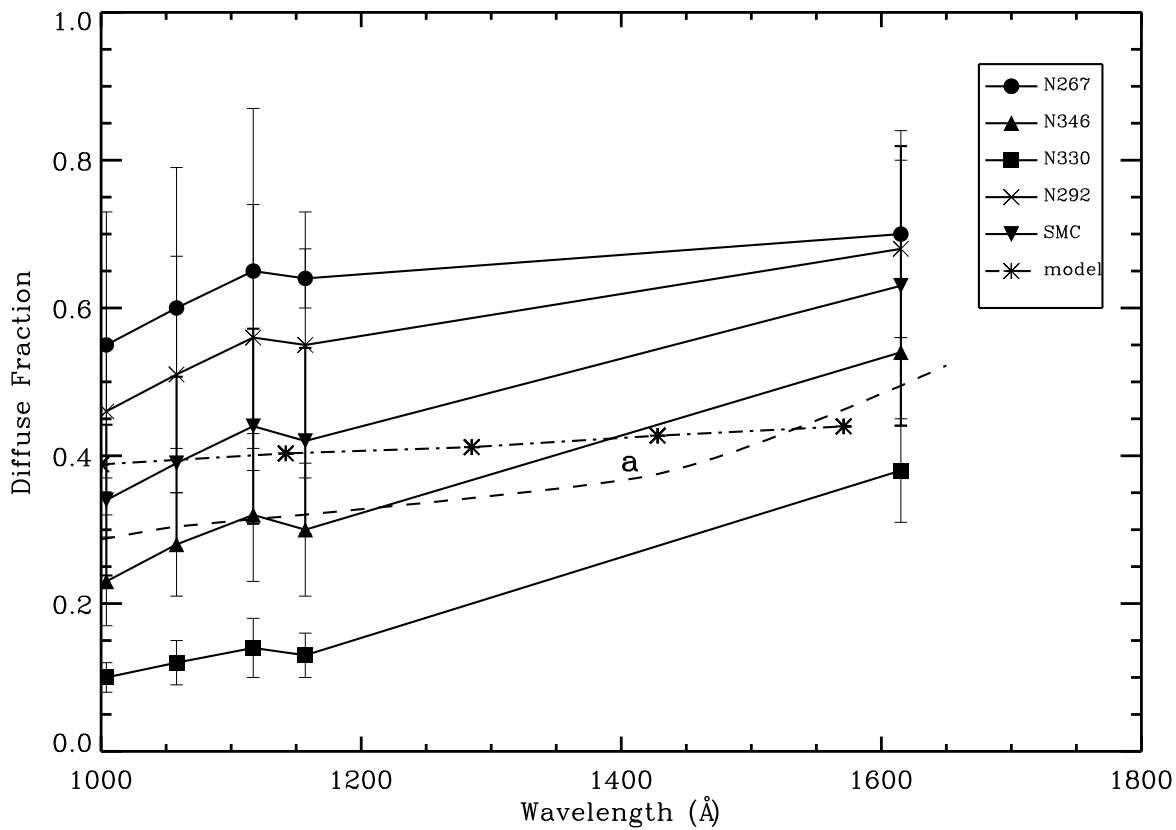}
\caption{Variation of diffuse fraction against the wavelength for the {\em UIT} regions as well as for the SMC bar as a whole. Dust albedo (dashed line) is from the model calculations of \citet{Weingartner01}. The error bars were empirically calculated by taking the extremes of the observed fluxes. The model calculation of diffuse fraction (dot-dashed line) is from \citet{Witt00}.
\label{Fig3}}
\end{figure}

The fraction of the total (stellar + diffuse) FUV light emitted as diffuse radiation in the SMC provides important information in context to the regional distribution of dust. We found the total flux in each of the {\em UIT} fields by summing the fluxes in all pixels in that field. We then used the catalog of \citet{Cornett97} to calculate the total stellar flux in each field. The diffuse flux in the {\em UIT} field was the difference between the two. We extended the stellar flux into the {\em FUSE} bands using Kurucz \citep{Kurucz92} model spectra and calculated their flux in {\em FUSE} bands. Finally, we extrapolated the diffuse flux into the {\em FUSE} bands using the observed {\em FUSE/UIT} diffuse flux ratios i.e., the slope of the best fit line (Figure \ref{Fig2}), obtained separately for each of the {\em FUSE} bands from their correlation with {\em UIT} band. \citet{Cornett97} predicted that 22\% of the diffuse flux was due to faint unresolved stars which we subtracted from each of the {\em UIT} and {\em FUSE} diffuse fluxes. The diffuse fraction defined as the diffuse emission divided by the total emission was then calculated for each region and over the entire SMC Bar (Figure \ref{Fig3}), with an estimated uncertainty of about 30\%. In all cases, the behavior of the diffuse fraction is almost the same, rising by 10\% from 1000 \AA\ to 1150 \AA\ and a further 50\% from 1150 \AA\ to 1615 \AA. The albedo of the dust obtained from the theoretical predictions of \citet{Weingartner01} for a mix of spherical carbonaceous and silicate grains increases by about the same factor over the considered wavelength range and the consequent increase in scattered light may be responsible for the increased diffuse fraction at longer wavelengths.

\begin{figure}
\plotone{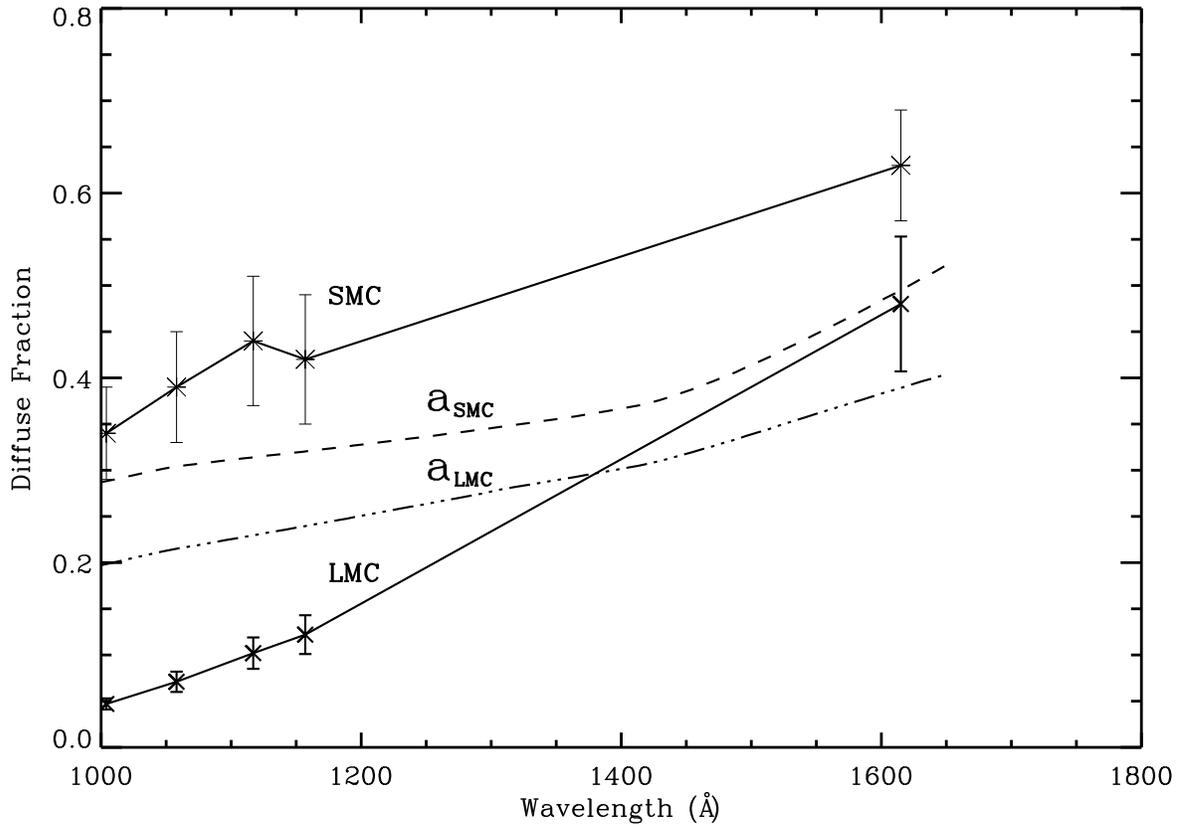}
\caption{Comparison of diffuse FUV fraction of the LMC and the SMC. Dashed line represents albedo of the SMC and the dot-dashed line represents the albedo of the LMC and are obtained from the model calculations of \citet{Weingartner01}.
\label{Fig4}}
\end{figure}

Integrating over the entire SMC Bar, we find that 34\% of the total radiation that escapes the SMC Bar at 1004 \AA\ is diffuse rising to 63\% at 1615 \AA. The scattered light in the SMC has been modeled by \citet{Witt00} using multiple scattering in a clumpy medium. They found that the diffuse radiation is 25\% to 50\% of the total (Figure \ref{Fig3}) depending on different dust geometries. Considering only H {\small II} regions of the SMC, we found that around 20\% of the total radiation at 1004\AA\ is diffuse rising to 50\% at 1615 \AA. Studies for the Orion nebula \citep{Bohlin82} and NGC 595 \citep{Malumuth96} find similar results with 66\% of the total radiation being diffuse at 1400 \AA\ in Orion and  55\% at 1700 \AA\ in NGC 595. \citet{Pradhan10} found significantly smaller values for the diffuse fraction in the LMC (Figure \ref{Fig4}) perhaps due to the difference in grain size and composition between the two galaxies \citep{Pei92,Weingartner01, Gordon03}. The albedo of the SMC dust is about 50\% higher \citep{Weingartner01} compared to the LMC dust (Figure \ref{Fig4}) and this may explain the increased diffuse fraction in the SMC.

We have examined the variation of the diffuse fraction over different region in the SMC bar finding that it is larger in those areas where there are fewer stars (NGC 267 and NGC 292) suggesting that much of the diffuse radiation from those regions is actually due to distant stars. Similar results were found in the LMC \citep{Pradhan10} which show that the diffuse fraction is less in crowded regions such as 30 Doradus, SN 1987A and N11 (4\% -- 10\%) and more in sparse regions such as N70 (24\% -- 45\%). \citet{Cole99} modeled the escape fraction of NUV photons for the LMC where they show that much of the stellar light is non-local i.e., the light from the distant OB associations is scattered by local dust. 

\begin{figure}
\plotone{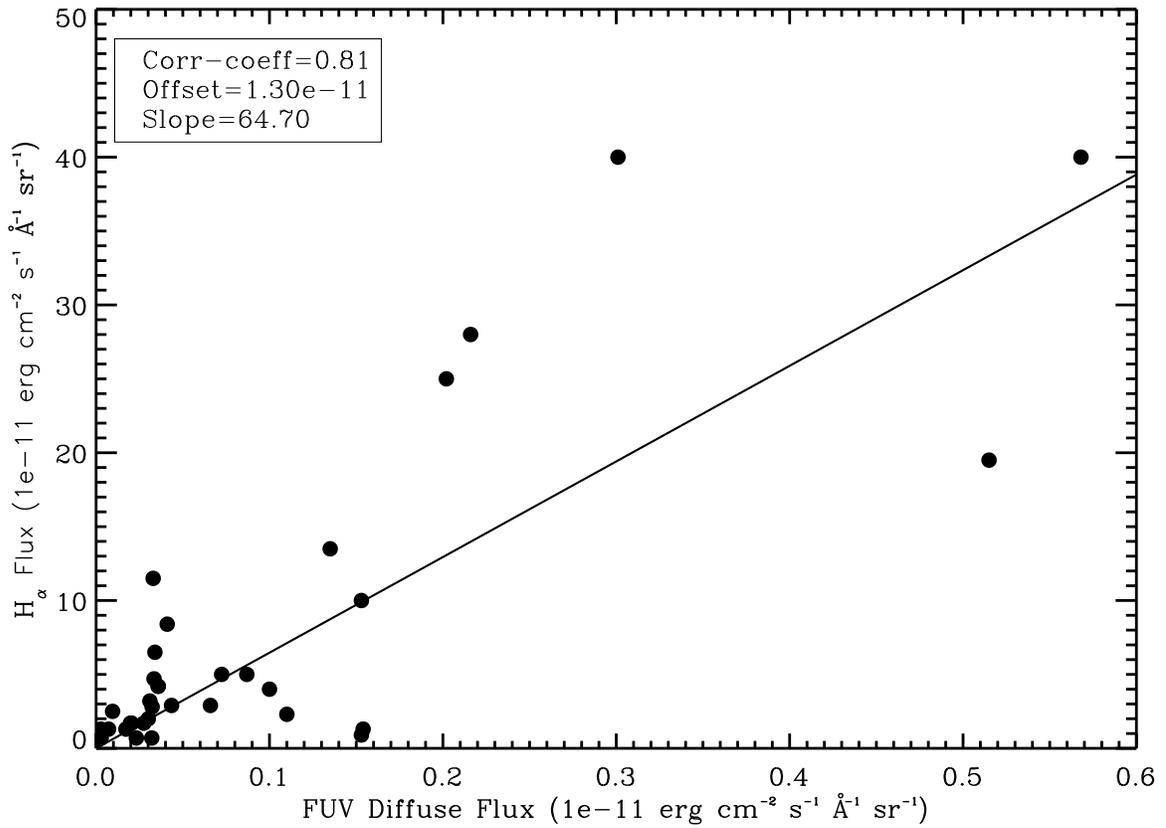}
\caption{Plot of the FUV diffuse flux and the H$\alpha$ flux of the H {\small II} regions of the SMC \citep{Kennicutt86}. The best fit line is with slope 64.70 and an offset of 1.30e-11 ergs cm$^{-2}$ s$^{-1}$ sr$^{-1}$ \AA$^{-1}$. The correlation coefficient is 0.81.
\label{Fig5}}
\end{figure}

A catalog of H {\small II} regions in the SMC was given by \citet{Davies76} and their integrated H$\alpha$ flux was calculated by \citet{Kennicutt86}. We have computed the integrated FUV diffuse flux in the {\em FUSE} bands for 36 H {\small II} regions defined by \citet{Cornett97}. We find a good correlation (r = 0.81) between the integrated diffuse FUV emission and H$\alpha$ emission from H {\small II} regions of the SMC (Figure \ref{Fig5}). This is as expected given that the H$\alpha$ flux is proportional to the brightness of the exciting stars as is FUV flux.

\section{Conclusions}

We have measured FUV diffuse emissions in the SMC using the spectra obtained by {\em FUSE} from different environments. The diffuse radiation is primarily due to light from hot stars scattered by the interstellar dust grains. We have used these observations to measure the FUV diffuse fraction which is 34\% -- 44\% in the {\em FUSE} bands (1000 -- 1150 \AA) increasing upto 63\% at 1615 \AA. The amount of light scattered increases towards the longer wavelengths showing that a large percent of the light at shorter wavelengths is absorbed by the dust. 

The behavior and distribution of FUV diffuse emission and emission fraction are quite similar in both the LMC and the SMC with much of the stellar radiation in both galaxies being non-local i.e., the diffuse (scattered) light in a particular region is the light coming from distant stars being scattered by local dust. The diffuse fraction in the SMC is higher than the LMC and the difference in diffuse fraction is related to amount of dust, dust grain properties, and geometry. A more detailed model incorporating the ample amount of data available for both galaxies in other spectral band is in progress to sort out the effect of local geometry from dust scattering. We found a good correlation between FUV diffuse emission and the H$\alpha$ emission in the H {\small II} regions.

AP acknowledges seed grant for research from Tezpur University. This research has made use of {\em FUSE} data. The {\it FUSE} was operated by Johns Hopkins University for NASA. We acknowledge the use of NASA Astrophysics Data System (ADS).

\newpage
\begin{deluxetable}{cccccccc}
\tabletypesize{\scriptsize}
\tablecaption{Details of the {\em FUSE} diffuse observations in the SMC. 
\label{tbl-1}}
\tablewidth{0pt}
\tablehead{
\colhead{{\em FUSE Id}} & \colhead{RA\tablenotemark{a}} & \colhead{Dec\tablenotemark{a}} & \colhead{LiF 1A1\tablenotemark{b}} & \colhead{LiF 1A2\tablenotemark{b}} & \colhead{LiF 1B1\tablenotemark{b}} & \colhead{LiF 1B2\tablenotemark{b}} & \colhead{{\it UIT}\tablenotemark{c}}
}
\startdata
G9310201	&00 46 38	&-73 08 24	&1.33	$\pm$ 0.83	&1.43	$\pm$ 0.36	&1.56	$\pm$ 0.45	&2.26	$\pm$ 0.53	& 16.02\\
G9310301	&00 47 16	&-73 08 24	& 0.97	$\pm$ 0.40	& 2.15	$\pm$ 0.22	& 2.56	$\pm$ 0.34	& 3.11	$\pm$ 0.38	& 32.96\\
G9310401	&00 48 26	&-73 19 12	& 2.78	$\pm$ 0.68	& 4.99	$\pm$ 0.29	& 6.62	$\pm$ 0.53	& 7.13	$\pm$ 0.65	& 24.69\\
G9310501	&00 49 02	&-73 14 24	& 3.28	$\pm$ 0.24	& 4.90	$\pm$ 0.13	& 5.78	$\pm$ 0.27	& 5.78	$\pm$ 0.24	& 18.90\\
G9310601	&00 51 07	&-73 21 36	& 9.07	$\pm$ 0.22	& 11.33	$\pm$ 0.11	& 11.09	$\pm$ 0.26	& 10.40	$\pm$ 0.28	& 13.57\\
F3230101	&00 53 07	&-74 39 00	& 3.87	$\pm$ 0.39	& 0.75	$\pm$ 0.63	& 0.48	$\pm$ 0.40	& 0.32	$\pm$ 0.29	& 4.19\\
F3230102	&00 53 07	&-74 39 00	& 0.15	$\pm$ 0.15	& 0.10	$\pm$ 0.09	& 0.28	$\pm$ 0.20	& 0.29	$\pm$ 0.23	& 7.84\\
F3230103	&00 53 07	&-74 39 00	& 0.87	$\pm$ 0.57	& 1.22	$\pm$ 0.40	& 0.94	$\pm$ 0.15	& 1.24	$\pm$ 0.15	& 11.25\\
F3230104	&00 53 07	&-74 39 00	& 0.12	$\pm$ 0.12	& 0.16	$\pm$ 0.11	& 0.27	$\pm$ 0.23	& 0.25	$\pm$ 0.14	& 7.61\\
F3230105	&00 53 07	&-74 39 00	& 0.15	$\pm$ 0.15	& 0.46	$\pm$ 0.32	& 0.40	$\pm$ 0.27	& 0.51	$\pm$ 0.27	&\\
F3230106	&00 53 07	&-74 39 00	& 0.25	$\pm$ 0.15	& 0.78	$\pm$ 0.06	& 0.64	$\pm$ 0.34	& 0.76	$\pm$ 0.23	&\\
D9110901	&00 53 57	&-70 37 48	& 0.22	$\pm$ 0.19	& 0.46	$\pm$ 0.41	& 0.22	$\pm$ 0.19	& 0.46	$\pm$ 0.41	&\\
G9310701	&00 58 19	&-72 17 24	& 10.07	$\pm$ 0.17	& 12.81	$\pm$ 0.10	& 13.69 $\pm$ 0.23	& 12.45	$\pm$ 0.22	&\\
P2030201	&00 59 36	&-72 07 48	& 17.03	$\pm$ 0.24	& 20.47	$\pm$ 0.17	& 25.15	$\pm$ 0.35	& 23.28	$\pm$ 0.36	&\\
C1580101	&00 59 43	&-72 09 36	& 3.50	$\pm$ 0.25	& 4.80	$\pm$ 0.13	& 5.72	$\pm$ 0.24	& 5.81	$\pm$ 0.30	&\\
S4057101	&01 00 09	&-72 08 24	& 8.96	$\pm$ 0.13	& 10.69	$\pm$ 0.07	& 9.63	$\pm$ 0.21	& 9.46	$\pm$ 0.18	&\\
G9310801	&01 00 24	&-71 33 36	& 5.78	$\pm$ 0.24	& 7.76	$\pm$ 0.16	& 9.35	$\pm$ 0.27	& 7.98	$\pm$ 0.34	&\\
G9310901	&01 03 16	&-72 09 36	& 15.90	$\pm$ 0.23	& 20.27	$\pm$ 0.12	& 22.02	$\pm$ 0.24	& 19.86	$\pm$ 0.22	&\\
G9311002	&01 03 33	&-72 02 24	& 20.30	$\pm$ 0.27	& 24.25	$\pm$ 0.14	& 25.15	$\pm$ 0.22	& 23.29	$\pm$ 0.28	&\\
C0830201	&01 03 36	&-71 58 48	& 14.64	$\pm$ 0.39	& 18.29	$\pm$ 0.17	& 16.64	$\pm$ 0.41	& 18.24	$\pm$ 0.41	&\\
G0350101	&01 03 48	&-71 58 12	& 5.31	$\pm$ 0.22	& 6.79	$\pm$ 0.12	& 7.67	$\pm$ 0.21	& 7.48	$\pm$ 0.25	&\\
E5110802	&01 03 52	&-72 54 00	& 2.69	$\pm$ 0.46	& 3.77	$\pm$ 0.20	& 4.41	$\pm$ 0.40	& 4.57	$\pm$ 0.44	&\\
F3210103	&01 03 52	&-72 07 48	& 17.09	$\pm$ 0.12	& 24.79	$\pm$ 0.07	& 34.14	$\pm$ 0.15	& 33.34	$\pm$ 0.14	&\\
F3210104	&01 03 52	&-72 07 48	& 22.18	$\pm$ 0.25	& 33.53	$\pm$ 0.11	& 46.05	$\pm$ 0.17	& 45.32	$\pm$ 0.21	&\\
E5110801	&01 03 57	&-72 54 36	& 0.70	$\pm$ 0.52	& 0.86	$\pm$ 0.45	& 0.92	$\pm$ 0.40	& 0.90	$\pm$ 0.50	&\\
A0750204	&01 04 00	&-72 01 48	& 24.65	$\pm$ 0.30	& 34.22	$\pm$ 0.16	& 30.03	$\pm$ 0.28	& 28.92	$\pm$ 0.27	&\\
C0830302	&01 04 33	&-71 59 24	& 2.97	$\pm$ 0.24	& 3.31	$\pm$ 0.17	& 2.92	$\pm$ 0.18	& 2.78	$\pm$ 0.19	&\\
G0350301	&01 04 48	&-72 01 12	& 8.17	$\pm$ 0.31	& 10.21	$\pm$ 0.17	& 9.93	$\pm$ 0.26	& 9.34	$\pm$ 0.40	&\\
D9044301	&01 05 12	&-72 23 24	& 0.95	$\pm$ 0.17	& 1.32	$\pm$ 0.08	& 0.55	$\pm$ 0.09	& 0.71	$\pm$ 0.12	&\\
D9044401	&01 06 19	&-72 05 24	& 1.85	$\pm$ 1.85	& 0.88	$\pm$ 0.27	& 1.64	$\pm$1.20	& 0.78	$\pm$ 0.57	&\\
\enddata
\tablenotetext{a}{RA \& Dec represent the {\em FUSE} LWRS position. Units of right ascension are hours, minutes, and seconds; units of declination are in degrees, arc minutes, and arc seconds.}
\tablenotetext{b}{The surface brightness of the diffuse radiation observed in the {\em FUSE} bands are in units of 10$^{4}$ photon cm$^{-2}$ s$^{-1}$ sr$^{-1}$ \AA$^{-1}$ and the uncertainties are 1$\sigma$ error bar.}
\tablenotetext{c}{{\em UIT} surface brightness in units of 10$^{4}$ photon cm$^{-2}$ s$^{-1}$ sr$^{-1}$ \AA$^{-1}$ and the error in the data is around 15\% \citep{Cornett97}.}

\end{deluxetable}

\newpage

\newpage

\newpage
\newpage

\newpage
\end{document}